\documentclass{article}

\usepackage[margin=1in]{geometry}
\usepackage{hyperref}
\usepackage{amsmath, amsfonts}
\usepackage{graphicx}
\usepackage{color}   
\usepackage{tikz}     
\usepackage{subcaption}
\usepackage{float}   
\usepackage{mathtools}
\usepackage[english]{babel} 
\usetikzlibrary{arrows.meta, calc, shapes.misc}

\newcommand{\sech}{\mathrm{sech}}

\renewcommand{\i}{\mathrm{i}}
\newcommand{\e}{\mathrm{e}}
\newcommand{\eps}{\varepsilon}

\newcommand{\diff}[2]{\frac{d #1}{d #2}}

\title{Borel-Pad\'{e} exponential asymptotics for the discrete nonlinear Schr{ö}dinger model with next-to-nearest neighbour interactions}
\author{Christopher J. Lustri$^1$\footnote{Corresponding Author. Electronic address: christopher.lustri@mq.edu.au}, In\^{e}s Aniceto$^2$, and P. G. Kevrekidis$^3$}
\date{
$^1$School of Mathematics and Statistics, University of Sydney, Camperdown 2050, Australia\\[2ex]
$^2$School of Mathematical Sciences, University of Southampton, Southampton SO17 1BJ, United Kingdom \\[2ex]
$^3$Department of Mathematics and Statistics, University of Massachusetts Amherst, Amherst, Massachusetts 01003; and Department of Physics, University of Massachusetts Amherst, Amherst, Massachusetts 01003 
}

\begin{document}

\maketitle

\begin{abstract}
In the present work we study discrete nonlinear
Schr{\"o}dinger models combining nearest (NN) and
next-nearest (NNN) neighbor interactions, motivated
by experiments in waveguide arrays. While we consider
the more experimentally accessible case of positive ratio $\mu$ of
NNN to NN interactions, we focus on the intriguing
case of competing such interactions $(\mu<0)$, where 
stationary states can exist only for $-1/4 < \mu < 0$. We analyze the
key eigenvalues for the stability of the pulse-like
stationary (ground) states, and find that such
modes depend exponentially on the coupling parameter
$\eps$, with suitable polynomial prefactors and
corrections that we analyze in detail. Very good agreement
of the resulting predictions is found with systematic
numerical computations of the associated eigenvalues. This analysis uses Borel-Pad\'{e} exponential asymptotics to determine Stokes multipliers in the solution; these multipliers cannot be obtained using standard matched asymptotic expansion approaches as they are hidden beyond all asymptotic orders, even near singular points. By using Borel-Pad\'{e} methods near the singularity, we construct a general asymptotic template for studying parametric problems which require the calculation of subdominant Stokes multipliers.
\end{abstract}

\section{Introduction}

The discrete nonlinear Schr{\"o}dinger (DNLS) model~\cite{chriseil,kev09} has, for numerous decades
now, constituted one of the most canonical examples
of Hamiltonian nonlinear lattice dynamical models.
This is not only because it combines the quintessential
features of lattice diffraction and nonlinearity, leading to a large number of intriguing localized
structures, but also because it arises in a 
wide range of physical applications. Arguably, the 
most central among these are optical waveguide
arrays, where many of the relevant ideas have been
theoretically proposed and experimentally 
demonstrated~\cite{cole,LEDERER20081}. Over the
past two decades, the area of atomic Bose-Einstein
condensates (BECs) in deep optical lattices has also contributed relevant
theoretical~\cite{brazhnyi} and experimental~\cite{RevModPhys.78.179} applications of DNLS models.
Indeed, very recently clear experimental realizations
of single- and multi-site solitary waves were reported
in this setting~\cite{cruickshank2025singlesitemultisitesolitonsbright}.
Additionally, the DNLS arises as a canonical ``envelope''
model for numerous other dispersive nonlinear settings
including granular crystals~\cite{Chong2018} and 
coupled lattices of electrical circuits~\cite{remoissenet}.

It is well-known that the standard DNLS model involves
a coupling of each node with its nearest neighbors (NN)
(as well as on-site nonlinearity)~\cite{chriseil,kev09}.
However, an important variant to the model was 
proposed in the work of~\cite{Efremidis2002}, where
a zigzag waveguide array was devised as a means for
{\it controllably} introducing (geometric, via suitable
overlap integrals) interactions with next-nearest-neighbors
(NNN)
in the system. Indeed, in that work, it was theoretically
argued that the relevant ratio of the coupling between
NNN and NN interactions could be practically 
varied widely, e.g.,
in a range from $0$ to approximately $4.5$, through a
variation of the angle between 3 neighboring waveguides
within the array. Importantly, this theoretical
proposal was experimentally implemented in 
femtosecond-laser-written waveguide arrays only
a few years later~\cite{Dreisow:08}. This remarkable result fueled significant further experimental
activity that  cemented this platform as
an excellent testbed for producing solitary waves
while exploring the NNN to NN interaction interplay~\cite{Szameit:09} and  fundamental
lattice dynamical ideas therein, such as the 
potential for Bloch oscillations~\cite{Dreisow:11}.
More recently this has, in turn, led to further 
theoretical developments such as the examination of
discrete solitary waves~\cite{Rothos2025}, including 
in the presence of {\it competing}
NN and NNN interactions, a topic of particular 
interest in its own right~\cite{HU2020126448}.

One of the key features that arise as a result
of the breaking of translational invariance 
inherent in the DNLS  model is
the existence of multiple solitary wave
solutions pertaining to the same
continuum limit soliton~\cite{kev09}. Notably,
there exists an on-site and an inter-site waveform,
the energy difference between which represents the
celebrated (in dislocation theory) Peierls-Nabarro
barrier~\cite{campbell}. This symmetry breaking
has been systematically observed to be associated
with {\it spectral} implications in the linearization
around the solitary waves~\cite{johaub,Todd_Kapitula_2001}.
Indeed, while the continuum, translationally invariant
soliton can (via Noether theory and conservation of
momentum) be associated with a neutral vanishing
pair of eigenvalues, this is not the case for the
discrete counterparts of the wave. The on-site waveform
features an imaginary pair of eigenvalues, while
the inter-site state features a real pair. 
Since the second difference can be approximated
by translationally invariant derivative operators
in a suitable Taylor expansion and all the latter
do not break the symmetry, it has been known for
some time that the relevant eigenvalue should
be exponentially dependent on the lattice 
spacing. This was originally approximated
with the right power law dependence (on the lattice
spacing) in front of the exponential, yet not
with the correct numerical prefactor~\cite{Todd_Kapitula_2001}. The recent
work of a subset of the present authors has amended this~\cite{lustri2025exponential},
and also determined the correction to the leading 
order approximation.

The analysis of the NN problem in \cite{lustri2025exponential} used factorial-over-power exponential asymptotics, developed in \cite{Chapman,Daalhuis}, to calculate the form of the exponentially-small eigenvalues in the NN problem, which lie beyond all orders of the standard series expansion. These methods match the divergent form of the asymptotic series terms with an inner expansion near complex singularities of the leading-order solution to compute the Stokes multipliers. In the NNN problem, however, the relevant inner expansion contributions are also beyond all orders \textit{within} the inner region, meaning that the Stokes multipliers cannot be determined using matched asymptotic expansion methods. Instead, we will use more powerful Borel transform methods to access beyond-all-orders contributions to the inner expansion.

It is well established that the large order behaviour of an asymptotic, factorially divergent expansion encodes the existence of {all} relevant (and irrelevant) exponential corrections \cite{aniceto2019primer}. These contributions are most evident in the so-called Borel plane. If the original asymptotic series is identified as the Laplace transform of an analytic function on some Riemann surface then the Borel transform gives its corresponding convergent expansion, which can generally be analytically continued to any region of the complex plane. The Borel transform contains branch point singularities corresponding to each of the exponentially dominant and subdominant contributions. The transformed result may therefore be studied near these singularities using complex analysis techniques such as analytic continuation of truncated convergent series, allowing us to find the overall strength of the corresponding exponential correction.

Pad\'{e} approximants \cite{graves1981pade} are a very efficient method of analytic continuation of a truncated convergent series, which approximate it as a rational function of two polynomials. Its singularities correspond to zeros of the denominator polynomial. In the case of branch cut singularities, these zeros will accumulate approximating the cut \cite{stahl1997convergence}. When applied to the Borel transform of an asymptotic series, the resulting function is called the Borel-Pad\'{e} approximant and allows us to calculate all exponential corrections, with the accuracy depending on the  number of series terms used. This method has been used successfully in many applications in theoretical physics, from string and field theoretic observables to kinetic theory describing strongly the evolution of coupled particle interactions 
 (for examples see \cite{aniceto2012resurgence, di2020looking, eynard2023arxiv,heller2022relativistic, marino2022new,serone2018lambdaphi4}).

The computation of the strength of branch point singularities of the Borel transform can be further simplified by using conformal maps to determine the so-called conformal Borel-Pad\'{e} approximant associated with a particular branch point of interest \cite{aniceto2019large,caliceti2007useful,costin2021conformal,jentschura2001improved}. This allows us to transform the original branch point singularity into a simple pole and determine the strength via a residue calculation. 

In the present work, we examine the intersection of
the two above themes. Motivated by the wide 
applicability and experimental relevance of the NNN
model, we adapt our exponential asymptotics approach
to study the effects of competing or/and
cooperating NNN and NN interactions. We identify technical challenges that arise in such
settings and utilize the so-called Borel-Pad\'{e} analysis
towards the estimation of the relevant dominant
eigenvalue of linearization around both on-site and inter-site solitary waves. In the process, we clearly
explain their parametric regime of existence, as well
as their resulting stability characteristics.
In Section 2, we briefly provide the model setup,
while in section 3, we analyze the solution 
asymptotics. Section 4 describes the implementation
of the Borel-Pad{\'e} analysis, and section 5
provides our stability conclusions. Section 6 provides a comparison with numerical observations,
while section 7 summarizes our results and presents our conclusions and
some interesting directions for future study.

\section{Model Setup}

The 1D discretized nonlinear Schr\"{o}dinger (DNLS) equation with next-to-nearest-neighbour (NNN) interactions~\cite{Efremidis2002,kev09} is  
given by:
\begin{equation}
\frac{\mu}{\varepsilon^2}(Q_{n+2}-2Q_n+Q_{n-2})+ {\rm i} \dot{Q}_n + \Delta_2 Q_n + |Q_n|^2 Q_n = 0,\label{e:Qn}
\end{equation}
where the  overdot denotes differentiation in time, while the discrete Laplacian with spacing 
$\varepsilon$ reads:
\begin{equation}
\Delta_2 Q_n = \frac{Q_{n-1} - 2 Q_n + Q_{n+1}}{\eps^2}.\label{e:Delta2}
\end{equation}
This form is inspired by setups in the form of zigzag waveguide arrays~\cite{Efremidis2002,Szameit:09}. In this formulation, $\mu$ parameterises the strength of the NNN interaction. We impose $\mu \neq -1/4$, as this choice requires different asymptotic scaling assumptions.
In what follows, we will mostly examine the case of competing
NNN and NN interactions, i.e., $\mu<0$, although we will also consider
the ``cooperative'' setting of $\mu>0$.

A notable case occurs for $\mu = -1/16$. Setting $F_n= \sqrt{3}Q_n/{2}$ and $\tau = 4t/3$ gives
\begin{equation}
    \i \dot{F}_n + \tfrac{1}{\varepsilon^2}\left(-\tfrac{1}{12}F_{n-2} + \tfrac{4}{3}F_n - \tfrac{5}{2}F_n + \tfrac{4}{3}F_{n+1} - \tfrac{1}{12}F_n\right) + |F_n|^2 F_n = 0,
\end{equation}
which is the fourth-order central difference approximation in space to the continuous nonlinear Schr\"{o}dinger equation. 

\section{Series Solution}
Much of the analysis in this section is analogous to that of \cite{lustri2025exponential}, so we only outline key steps here. We use the well-known standing wave form $Q_n = {\rm e}^{{\rm i} t} q_n$ in \eqref{e:Qn} to obtain the stationary DNLS with NNN (and NN) interactions,
\begin{equation}
\frac{\mu}{\varepsilon^2}(q_{n+2}-2q_n+q_{n-2}) + \frac{1}{\varepsilon^2}({q_{n+1}-2q_n + q_{n-1}})+ (|q_n|^2 - 1)q_n  = 0.
\end{equation}
We take the continuum limit $z = \eps n$ and $q_n \equiv q(z)$, and expand as a power series in $\eps$ to obtain
\begin{equation}
   2\mu\sum_{m=1}^{\infty}\frac{2^{2m}\eps^{2m-2}}{(2m)!}\diff{^{2m}q(z)}{z^{2m}}+2\sum_{m=1}^{\infty}\frac{\eps^{2m-2}}{(2m)!}\diff{^{2m}q(z)}{z^{2m}} + (|q(z)|^2 -1)q(z) = 0.\label{eq:infode}
\end{equation}
We write $q(z)$ as an asymptotic power series in $\eps$ such that
\begin{equation}\label{e:series}
q(z) \sim \sum_{j=0}^{\infty} \eps^{2j}q_j(z) \quad \mathrm{as} \quad \eps \to 0.
\end{equation}
Substituting \eqref{e:series} into \eqref{eq:infode} and balancing powers of $\eps$ leads to leading order to
\begin{equation}
    \beta^2 \diff{^2q_0(z)}{z^2} + (|q_0(z)|^2-1)q_0(z) = 0,
\end{equation}
where $\beta = \sqrt{{{1+4\mu}}}$, with $\mu \neq 0$. This can be solved to give the leading-order standing wave
\begin{equation}\label{eq:q0}
    q_0(z) = \sqrt{2}  \,   \sech\left(\frac{\tilde{z}}{\beta}\right),
\end{equation}
where $\tilde{z} = z - z_0$ for some constant $z_0$ offset, and we have exploited the phase/gauge invariance of the model to select the real solution. The solution $q_0(z)$ is singular at points $\tilde{z} = \tilde{z}_s$, with asymptotic behavior
\begin{equation}\label{e:zsnnn}
 \tilde{z}_s = \pm\beta \left[\frac{\pi\mathrm{i}}{2} + k \pi \mathrm{i}\right], \qquad q_0(z) \sim -\frac{\mathrm{i}\beta \sqrt{2}}{\tilde{z}-\tilde{z}_s} + \frac{\i (\tilde{z} - \tilde{z}_s)}{3\beta\sqrt{2}}\quad \mathrm{as} \quad \tilde{z}\to\tilde{z}_s,
\end{equation}
where $k \in \mathbb{Z}$. On the real axis, the leading-order solution has the large-$\tilde{z}$ asymptotic behavior $q_0(z) \sim 2\sqrt{2}\e^{-\tilde{z}/\beta}.$ Substituting \eqref{e:series} into \eqref{eq:infode} and noting that $q_0$ is real and positive gives the recurrence relation
\begin{equation}\label{e:recur}
2\sum_{m=1}^{\infty} \frac{c_m}{(2m)!} \frac{d^{2m} q_{j-m+1} }{d\tilde{z}^{2m}}+ 2 q_0^2 |q_j |+ (q_0^2 -1)q_j = 0, \qquad c_m = 1 + 2^{2m}\mu.
\end{equation}
Using this recurrence relation with $q_0$, we obtain an equation for $q_1(\tilde{z})$,
\begin{equation}
    \beta^2 \frac{d^2 q_1}{d\tilde{z}^2} +(1 + 16\mu)\frac{1}{12}\frac{d^4 q_0}{d\tilde{z}^4} + (3q_0^2 - 1)q_1 = 0.
\end{equation}
Solving this exactly gives the particular solution with asymptotic behavior
\begin{align}\label{e:q1nearzs}
q_1(\tilde{z}) 
= \frac{(1+16\mu)(9\beta-7\beta\cosh(2\tilde{z}) + \tilde{z} \sinh(2\tilde{z}))}{24\beta^5\sqrt{2}\cosh^3(\tilde{z})}&\\
    \sim \frac{{\rm i}\sqrt{2}(1 + 16\mu)}{3\beta(\tilde{z}-\tilde{z}_s)^3} + \frac{\pi(1+16\mu)}{24\beta^2\sqrt{2}(\tilde{z}-\tilde{z}_s)^2}&\quad \mathrm{as} \quad \tilde{z} \to \tilde{z}_{s}.\nonumber
\end{align}

\subsection{Late-Order Asymptotics}\label{S:expasymp}
As in \cite{lustri2025exponential}, we expect that the solution will diverge in a factorial-over-power fashion, and hence follow the method of \cite{Chapman, Dingle}. We assume that the asymptotic behavior of $q_j$ as $j \to \infty$ can be written a sum of factorial-over-power terms with the form
\begin{equation}\label{e:ansatz}
q_j(\tilde{z}) \sim \frac{U(\tilde{z})\Gamma(2j + \gamma)}{\chi(\tilde{z})^{2j+\gamma}},
\end{equation}
where $\chi(z) = 0$ at some $\tilde{z} = \tilde{z}_s$, and each singular point $\tilde{z}_s$ generates late-order terms in the asymptotic behavior. We obtain equations for $\chi$ and $U$ by substituting the ansatz \eqref{e:ansatz} into the recurrence relation \eqref{e:recur}, and balancing terms of the same asymptotic order in the limit of $j \to \infty$.

\subsubsection{Calculating \texorpdfstring{$\chi$}{chi}}

Balancing terms of $\mathcal{O}(q_{j+1})$ as $j \to \infty$ gives
\begin{equation}
\label{e:singulanteqnnn}
2  U \sum_{m=1}^{j-1} \frac{c_m (\chi')^{2m}}{(2m)!}  = 0,
 \end{equation}   
where $c_m$ is defined in \eqref{e:recur} and a prime denotes differentiation with respect to $\tilde{z}$. We can find the leading-order solution as $j \to \infty$ by extending the upper bound of the summation term to infinity, finding
\begin{equation}
    \mu (\cosh(2\chi')-1)+(\cosh(\chi') - 1)=0.
\end{equation}
This gives the solutions for $\chi$ from the nearest neighbor analysis, $\chi' = 2\pi\mathrm{i}\ell$ for $\ell \in \mathbb{Z}$, as well as another family of solutions given by
\begin{align}\label{e:chifam}
    \chi' &= 2\pi{\rm i}\ell + \log\left[-\tfrac{1}{2\mu}{(1+2\mu\pm\beta)}\right], \quad \ell \in \mathbb{Z},
\end{align}
where each choice of $\ell$ and the sign gives a possible value for $\chi'$. The  late-order term asymptotic behavior \eqref{e:ansatz} is dominated by contributions with the smallest value of $|\chi'|$. For \eqref{e:chifam}, this corresponds to the two choices with $\ell = 0$. Consequently, for $\mu < 0$ we need only consider four contributions, which we denote as
\begin{align}
    \chi_{1,\pm}' = \pm2\pi\i,\qquad \chi_{2,\pm}' = \log\left[-\tfrac{1}{2\mu}{(1+2\mu\pm\beta)}\right].
\end{align}

For $\mu > 0$, \eqref{e:chifam} has four solutions with equal magnitude; we denote these where appropriate as $\chi_{2,\pm}'$, which have $\mathrm{Im}(\chi') = \pi$ and $\overline{\chi}_{2,\pm}'$, which have $\mathrm{Im}(\chi') = -\pi$. 

\begin{figure}
\centering
\subfloat[Values of $\chi'$ as $\mu$ varies]{
\includegraphics[height=0.2\textwidth]{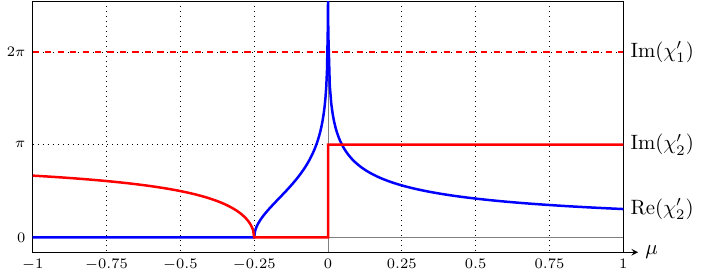}
}
\subfloat[Typical values of $\chi'$]{
\includegraphics[height=0.25\textwidth]{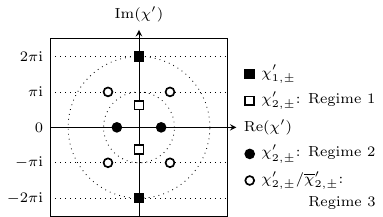}
}
\caption{(a) Real parts (blue) and imaginary parts (red) of $\chi_{1,+}'$ (dashed) and $\chi_{2,+}'$ (solid) as $\mu$ varies. As $\chi_{1,+}'$ is imaginary, its real part is not included in this figure. (b) Schematic of typical values for $\chi_{1,\pm}'$ and $\chi_{2,\pm}'$ in each regime. $\chi'_{1,\pm}$ is always equal to $2\pi\mathrm{i}$, while $\chi'_{2,\pm}$ varies depending on which regime is selected by the choice of $\mu$.}
\label{fig:nnn}
\end{figure}

The behavior of $\chi_{1,+}'$ and $\chi_{2,+}'$ is shown in Figure \ref{fig:nnn}(a). This figure depicts 3 regimes with distinct solution behaviour. 
In Regime 1 ($\mu < -1/4$), $\chi_{2,\pm}'$ are imaginary. In Regime 2 ($-1/4 < \mu < 0$), $\chi_{2,\pm}'$ are real. In Regime 3 ($\mu > 0$), $\chi_{2,\pm}'$ are complex with nonzero real and imaginary parts. 

In Regime 1, it is significant that $-\pi < \mathrm{Im}(\chi_{2,\pm}') < \pi$. Note also that $|\chi_{2,\pm}'| < |\chi_{1,\pm}'|$, or $2\pi$, for all but very small values of $\mu$ (ie. $-0.00186\ldots < \mu < 0.00437\ldots$). 

\subsubsection{Calculating \texorpdfstring {$U$}{U}}\label{S:Q}
At $\mathcal{O}(q_{j+1/2})$ as $j \to \infty$, we balance terms to obtain
\begin{equation}
 2 \mu \sum_{m=1}^{j-1} \frac{U' c_m (\chi')^{2m-1}}{(2m-1)!} = 0.
 \end{equation}
 Taking the sums to infinity, we obtain
 \begin{equation}
 2 U' \left(2 \mu \sinh(2\chi') + \sinh(\chi') \right) = 0.\label{eq:vaporiseQ}
 \end{equation}
As in \cite{lustri2025exponential}, when $\chi' = \chi_{1,\pm}'$ this statement is always true when $j \to \infty$ irrespective of $U$. We also find that this cancellation does not occur for $\chi_{2,\pm}'$, and therefore $U' = 0$, and hence $U$ is a constant, for exponential terms associated with $\chi_{2,\pm}$.

To obtain the expression for $U$ associated with $\chi_{1,\pm}$, we balance terms at the next order as $j \to \infty$, $\mathcal{O}(q_j)$, to find
 \begin{align}
\sum_{m=1}^{j-1} \frac{U'' c_m (\chi')^{2m-2}}{(2m-2)!} + (3 q_0^2  - 1 )U  &= 0.\label{e:UV.1}
 \end{align}
Extending the summation term to infinity gives, after some algebraic manipulation, 
\begin{equation}\label{e:Uode}
\beta^2 U'' + \left(6     \,  \sech^2\left({\tilde{z}}/{\beta}\right) - 1\right) U  = 0.
\end{equation}
This is identical to the equivalent equation in \cite{lustri2025exponential} up to an appropriate scaling of $U$ and $\tilde{z}$, giving $U(\tilde{z}) = K_1 U_1(\tilde{z}) + K_2 U_2(\tilde{z})$
,where $K_1$ and $K_2$ are constants that have yet to be determined, and
\begin{align}
U_1(\tilde{z}) = \tfrac{1}{2} \tanh\left(\tfrac{\tilde{z}}{\beta}\right)\sech\left(\tfrac{\tilde{z}}{\beta}\right),
\quad U_2(\tilde{z}) = \sech\left(\tfrac{\tilde{z}}{\beta}\right)     \left(\tfrac{6\tilde{z}}{\beta}\tanh\left(\tfrac{\tilde{z}}{\beta}\right)+ \cosh\left(\tfrac{2\tilde{z}}{\beta}\right)-5\right).\label{e:U2}
\end{align}
 The limiting behavior as $\tilde{z} \to \infty$ of these solutions is $U_1(\tilde{z}) \sim e^{-\tilde{z}/\beta}$ and $U_2(\tilde{z}) \sim e^{\tilde{z}/\beta}$. The asymptotic behavior of these solutions as $\tilde{z} \to \tilde{z}_s$ is, 
\begin{align}
U_1(\tilde{z}) \sim -\frac{\rm i \beta^2}{2(\tilde{z} - \tilde{z}_s)^2},\qquad U_2(z) \sim \frac{3\pi \beta^2}{(\tilde{z}-\tilde{z}_s)^2}.\label{e:Uasymp}
\end{align}

\subsubsection{Calculating \texorpdfstring{$\gamma$}{gamma}}

For the late-order terms associated with $\chi_{1,\pm}$, we follow \cite{lustri2025exponential} and use use the local behavior of $q_0$ in \eqref{e:Uode} to find that the asymptotic behavior of the solution near $\tilde{z} = \tilde{z}_s$ is
\begin{equation}\label{e:localU}
U(\tilde{z}) \sim \Lambda_1 (\tilde{z}-\tilde{z}_s)^3 + \frac{\Lambda_2}{(\tilde{z}-\tilde{z}_s)^2} \quad \mathrm{as} \quad \tilde{z} \to \tilde{z}_s,
\end{equation}
where $\Lambda_1$ and $\Lambda_2$ are arbitrary constants yet to be determined. The asymptotic behavior of $q_j$ near the singularity is dominated as $j \to \infty$ by the term containing $\Lambda_1$. Setting $U \sim \Lambda_1 (z-z_s)^3$ near the singular point gives the local behavior
\begin{equation}\label{e:localansatz}
q_j \sim \frac{\Lambda_1(\tilde{z}-\tilde{z}_s)^3 \Gamma(2j + \gamma)}{[2\pi {\rm i}(\tilde{z} - \tilde{z}_s)]^{2j+\gamma}}\quad \mathrm{as} \quad j\to\infty,     \tilde{z}\to \tilde{z}_s.
\end{equation}
In order to be consistent with the leading-order asymptotics for $q_0$, we require that the singularity have strength of 1 when $j = 0$. This gives $\gamma = 4$. 

The constants $\Lambda_1$ and $\Lambda_2$ may be used to find the constants $K_1$ and $K_2$. Expanding the $U$ from \eqref{e:Uasymp} about the point $\tilde{z} = \tilde{z}_s$ and comparing the result with \eqref{e:localU} gives $\Lambda_1 = {4{\rm i}}K_2/5\beta^3$ and $ \Lambda_2 = -{\rm i}K_1/2 + 3 \pi K_2$. Hence the asymptotic behavior of $q_j$ as $j \to \infty$ is dominated by factorial-over-power terms with the form
\begin{equation}
q_j \sim \frac{5\beta^3 \Lambda_1}{4 {\rm i}}\frac{ U_2(z) \Gamma(2j + 4)}{ [2\pi {\rm i}(\tilde{z}-\tilde{z}_s)^{2j+4}]} \quad \mathrm{as} \quad j \to \infty.
\end{equation}
In subsequent analysis, we will omit the subscript on $\Lambda_1$ and $U_2$. To emphasise that $\Lambda$ depends on the choice of $\mu$, we will write it as $\Lambda(\mu)$. 

For the late-order terms associated with $\chi_{2,\pm}$, it is straightforward to show that $\gamma = 1$, as $U$ is constant. We will not require this result in any subsequent analysis.

At this stage, we would typically calculate $\Lambda(\mu)$, but this is not straightforward for the NNN analysis. To calculate $\Lambda(\mu)$, we observe that the asymptotic series \eqref{e:series} with late-order term behavior \eqref{e:ansatz} fails to be asymptotic in a region of width $\mathcal{O}(\eps)$ around $\tilde{z} = \tilde{z}_s$. We then apply the rescaling $\eps\eta = \tilde{z} - \tilde{z}_s$ and $\hat{q}(\eta) = q(\tilde{z})/\eps$. At leading order in $\eps$, the inner equation is given by
\begin{align}
2\sum_{m=1}^{\infty} \frac{c_m}{(2m)!}\frac{d^{2m}\hat{q}(\eta)}{d\eta^{2m}} + |\hat{q}(\eta)|^2\hat{q}(\eta) &= 0.\label{e:innereq3}
\end{align}
The matched asymptotic expansion approach requires expanding $\hat{q}(\eta)$ as a power series as $|\eta| \to \infty$, and using Van Dyke's matching principle in this limit to determine the value of $\Lambda(\mu)$ \cite{Chapman}. This approach was used in \cite{lustri2023locating}; however, it is not possible to use this method to determine values of $\Lambda(\mu)$ associated with $\chi_{1,\pm}$. 

As $|\chi_{2,\pm}'| < |\chi_{1,\pm}'|$ for most of the region of interest, the late-order behavior is factorially dominated by terms of the form \eqref{e:ansatz} and $\chi = \chi_{2,\pm}$, and the terms associated with $\chi = \chi_{1,\pm}$ lie beyond all orders of the expansion as $|\eta| \to \infty$. 
This means that matched asymptotic expansions can be used to find the values of $\Lambda(\mu)$ associated with $\chi_{2,\pm}$ but not to find the values of $\Lambda(\mu)$ associated with $\chi_{1,\pm}$, which are required to calculate the eigenvalues in section \ref{S:Stability}. 

Instead, we will instead use a Borel-Pad\'{e} approach to study the Stokes switching within this inner region and compute exponentially small inner-region contributions. We will match the Stokes multipliers within this region to the late-order ansatz \eqref{e:localansatz} to calculate $\Lambda(\mu)$. We will describe these calculations in section \ref{S:BorelPade}.

\subsection{Stokes Phenomenon}\label{S:stokesphen}

As in \cite{lustri2025exponential}, the Stokes switching procedure is very similar to \cite{king_chapman_2001, Daalhuis}. The introduction of the NNN terms does alter the intermediate analysis slightly from that described in \cite{king_chapman_2001}, although the final result is identical. As the analysis is altered slightly, the main adapted steps are described in Appendix \ref{app:stokes}.

In this analysis we find that the quantity which appears on the right-hand side of the Stokes line associated with $\chi_{1,\pm}$, denoted $q_{\mathrm{exp}}$, is given by
\begin{equation}\label{e:remainder}
q_{\mathrm{exp}} \sim \frac{2\pi}{\eps^4} U e^{-\beta\pi^2/\eps}\sin\left(\frac{2\pi \tilde{z}}{\eps}\right) \quad \mathrm{as} \quad \eps \to 0.
\end{equation}
Unlike the nearest-neighbour analysis in \cite{lustri2025exponential} there is also a second exponential contribution present in the asymptotic solution, associated with $\chi_{2,\pm}$. As $\mathrm{Re}(\chi_{2,\pm}) \neq 0$, this contribution decays exponentially quickly as $|\tilde{z}| \to \infty$. For sufficiently large $\tilde{z}$, the oscillations in \eqref{e:remainder} are therefore the only remaining asymptotic contribution.

\subsection{Exponential Contributions}\label{S:expcont}

Imaginary values of $\chi'$ produce  exponentially small oscillations that grow as $\tilde{z} \to \infty$, such as those seen in the nearest-neighbor analysis. In general, exponential contributions to the solution have the form \cite{Chapman, Daalhuis}
\begin{equation}
    \mathcal{S} U e^{-\chi/\varepsilon} + \mathrm{c.c.},\label{eq:qexp0}
\end{equation}
where $\mathcal{S}$ is the Stokes multiplier, and c.c. denotes the complex conjugate term. This form was used to determine $q_{\mathrm{exp}}$ in \eqref{e:remainder}. Stokes lines in the system satisfy the condition $\mathrm{Im}(\chi) = 0$ and $\mathrm{Re}(\chi) = 0$. Exponentially-small asymptotic contributions appear in the solution as these Stokes lines are crossed.

\begin{figure}
\centering
\subfloat[$\chi_{1,\pm}$: All regimes]{
\includegraphics[width=0.31\textwidth]{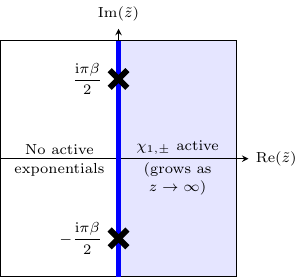}
}
\subfloat[$\chi_{2,\pm}$: Regime 1]{
\includegraphics[width=0.31\textwidth]{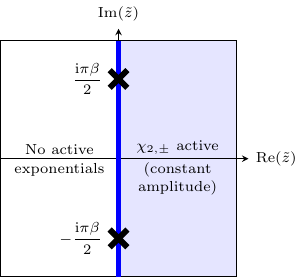}
}
\subfloat[$\chi_{2,\pm}$: Regime 2]{
\includegraphics[width=0.31\textwidth]{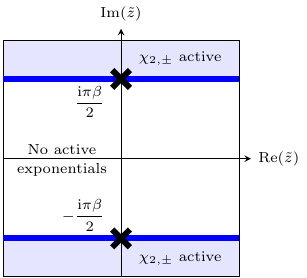}
}
\caption{Stokes structure for the exponential contributions associated with (a) $\chi_{1,\pm}$ and (b)--(c) $\chi_{2,\pm}$ in Regime 2 and 3. Crosses represent leading-order singularities that generate Stokes lines. Stokes lines are shown as solid blue lines. Shaded regions contain at least one exponential contribution, and unshaded regions contain none. In (a), there is an exponential contribution to the right of the Stokes line that grows large as $z \to \infty$. In (b) there is a constant-amplitude exponential contribution to the right of the Stokes line. In (c) there are no exponential contributions for real $\tilde{z}$.}\label{fig:stokes}
\end{figure}

Imaginary values of $\chi'$ correspond to the appearance of oscillatory contributions when $z \in \mathbb{R}$ whose exponential factor does not grow or decay. This includes the contribution $q_{\mathrm{exp}}$ from \eqref{e:remainder}. Complex or real values of $\chi'$ correspond to the appearance of contributions that decay exponentially as $|\tilde{z}| \to \infty$. The Stokes structure for $\chi_1$ is shown in Figure \ref{fig:stokes}(a), and for $\chi_2$ in Figure \ref{fig:stokes}(b)--(c) for Regimes 1 and 2 respectively.

For all $\mu$, $\chi_1'$ is imaginary. Hence the associated exponential contributions in \eqref{e:remainder} are oscillatory on the real axis, and these oscillations appear as the Stokes lines on the imaginary axis are crossed from left to right. Since $U$ grows exponentially as $z \to \infty$, the oscillation amplitude grows in the limit that $z$ becomes large.

In Regime 1, exponential contributions associated with $\chi_{2,\pm}$ are present in the solution for $\mathrm{Re}(\tilde{z}) > 0$. As $\chi'_{2,\pm}$ are imaginary and $U$ is constant, the form of \eqref{eq:qexp0} indicates that the associated oscillations have constant amplitude as $|\tilde{z}| \to \infty$. 

In Regime 2, $\chi'_{2,\pm}$ is real, and the exponential contributions associated with $\chi_{2,\pm}$ are not present for $\tilde{z} \in \mathbb{R}$. We must still consider these contributions in the local analysis near $\tilde{z} = \tilde{z}_s$ to determine $\Lambda(\mu)$. Regime 3 (not shown) possesses exponentially small contributions on the real line that decay as $|\tilde{z}| \to \infty$.

\subsection{Site selection}

For $\mu > -1/4$ (Regimes 2 and 3), we see similar behavior to the NN analysis in \cite{lustri2025exponential}. 
The oscillations in \eqref{e:remainder} associated with $\chi_{1,\pm}$ are exponentially small when they switch in amplitude across the Stokes line at $\tilde{z} = 0$; however, because $U \sim K_2 \e^{\tilde{z}/\beta}$ as $\tilde{z} \to \infty$, the oscillations in \eqref{e:remainder} grow exponentially in this limit. This is not physically reasonable, so solutions can only exist if these oscillations are cancelled out entirely.

As $\tilde{z} \to \infty$ and $\eps \to 0$, the remainder satisfies the expression
\begin{equation}\label{e:remainder0}
q_{\mathrm{exp}} \sim \frac{2\pi}{\eps^4} \left(\frac{5\beta^3\Lambda(\mu)}{4{\rm i}}  e^{\tilde{z}/\beta}\right) e^{-\beta \pi^2/\eps}\sin\left(\frac{2\pi \tilde{z}}{\eps}\right).
\end{equation}
We can write this in the original discrete scaling using $\tilde{z} = \eps (n - n_0)$, and using the fact that $\Lambda(\mu)$ is a negative imaginary number, to obtain
\begin{equation}\label{e:remainder1}
q_{\mathrm{exp}} \sim -\frac{5\beta^3\pi|\Lambda(\mu)|}{2\eps^4}   e^{\eps(n-n_0)/\beta} e^{-\beta\pi^2/\eps}\sin\left(2\pi (n-n_0)\right)\quad \mathrm{as}\quad n \to \infty,     \eps \to 0.
\end{equation}
The presence of a slowly growing oscillatory tail in the asymptotic solution indicates that this solution can only exist if $n_0$ is chosen in a way such as to eliminate the oscillations entirely for integer values of $n$. There are only two values in $n_0 \in [0,1)$ that achieve this; $n_0 = 0$, or on-site pinning, and $n_0 = 1/2$, or inter-site pinning. Hence, even when the system contains NNN interactions, the only standing wave solutions to \eqref{e:Qn} are site-centered and inter-site-centered solutions.

For $\mu < -1/4$, the solution contains two oscillating contributions with different phases as $|z| \to \infty$. The oscillatory contribution associated with $\chi_{1,\pm}$ can be eliminated by selecting an on-site ($n_0 = 0$) or inter-site ($n_0 = 1/2$) configuration. From \eqref{eq:qexp0}, the exponential contribution associated with $\chi_{2,\pm}$ is proportional to $\sin\left(\omega(n - n_0)\right)$, where $\omega = \mathrm{Im}(\chi_{2,+}')$, 
and where $n_0$ has been chosen to obtain on-site or inter-site pinning. These oscillations can only be eliminated for large $\tilde{z}$ if $\omega$ is a multiple of $\pi$; however, this is never true for $\mu < -1/4$ because $0 < \omega < \pi$ (see Figure \ref{fig:nnn}). Consequently, the model does not possess standing wave solutions if $\mu < -1/4$.

\section{Borel-Pad\'{e} Analysis}\label{S:BorelPade}

To determine the Stokes multipliers, we need to expand solutions to the inner equation \eqref{e:innereq3} in the limit $|\eta| \to \infty$. This is a non-parameteric asymptotic expansion, as the asymptotic limit is in terms of the independent variable $\eta$, rather than a parametric asymptotic expansion (such as the limit $\eps \to 0$). That means we can employ the process from \cite{aniceto2019large} to determine $\Lambda(\mu)$, which is associated with exponentially subdominant contributions in the inner region.

We begin with the inner equation \eqref{e:innereq3}, corresponding to the local solution in the neighbourhood of the singularity, and perform the following steps:
\begin{enumerate}
 \item Find a transseries solution for $\hat{q}(\eta)$ obeying \eqref{e:innereq3}, formally including any exponential corrections that appear in the solution.
 \item Take the Borel transform of the algebraic series in $\eta$, and analyse the result using Pad\'{e} approximants to locate branch points in the Borel plane.
 \item Adjust the Borel transform parameter and apply a conformal map in the Borel plane to replace the key logarithmic branch point by a simple pole.
 \item Compute a Pad\'{e} approximant for the mapped function and numerically calculate the residue of the simple pole. This residue gives us direct access to the Stokes constants, which match to $\Lambda(\mu)$ in the original problem.
 \end{enumerate}

 \subsection{Transseries solution}

The general solution to \eqref{e:innereq3} can be written formally in the regime of large variable $\eta\gg1$ as a transseries,
\begin{equation}
\hat{q}(\eta;\sigma_{i})=\sum_{i=1}^{N}\sum_{m_{i}\ge 0}\prod_{j}\left(\sigma_{j}\mathrm{e}^{-A_{j}\eta}\right)^{m_{j}}\Phi^{(\mathbf{m})}(\eta),\quad\mathbf{m}\equiv\left(m_{1},\cdots,m_{N}\right),\label{e:transansatz}
\end{equation}
which is a series of exponential terms with $N$ singulants $A_{i}$ (where $N$ can be infinite, as it is here), and $\Phi^{(\mathbf{m})}(\eta)$ are Laurent series in inverse powers of $\eta$. As the equation is nonlinear, we must include all powers and products of the exponential terms, indexed by $\mathbf{m}$. The parameters $\sigma_{i}$ are fixed by boundary conditions. For the current analysis we only require the leading exponential contributions for each relevant singulant, 
\begin{equation}
\hat{q}(\eta,\sigma_{i})=\Phi^{(0)}(\eta)+\sum_{i}\sigma_{i}     \mathrm{e}^{-A_{i}\eta}     \Phi^{(i)}\left(\eta\right)+\mathrm{higher\:exponential\:orders}.\label{eq:simpler-ansatz}
\end{equation}
The singulants $A_{i}$ and the Laurent series $\Phi^{(0)}(\eta)$ can be determined by applying the ansatz \eqref{e:transansatz} to \eqref{e:innereq3}. We first write the Laurent series
\begin{equation}
\Phi^{(i)}(\eta)=\sum_{\ell=0}^{\infty}\frac{a_{\ell}^{(i)}}{\eta^{\ell + \nu_i}}  .\label{eq:transs-pert-ansatz}
\end{equation}
To obtain an appropriate asymptotic balance in \eqref{e:innereq3} as $|\eta| \to \infty$ we require the expansion to contain powers of $\eta^{-1-2\ell}$ and $\nu_0=1$. Balancing powers of $\eta$ shows that odd coefficients vanish, or $a_{2\ell+1}^{(0)}=0$, and even coefficients obey $a_{0}^{(0)} =-\mathrm{i}\sqrt{2}\beta$ and
\begin{align}
((2\ell+1)(2\ell+2)-6)&c_{1}     a_{2\ell}^{(0)}  =-\sum_{n=1}^{\ell}c_{n+1}     a_{2\ell-2n}^{(0)}\frac{\Gamma(3+2\ell)}{\Gamma(1+2\ell-2n)}\label{eq:recurborel} \\
 & -3a_{0}^{(0)}\sum_{\ell_{2}=1}^{\ell-1}a_{2\ell_{2}}^{(0)}     a_{2\ell-2\ell_{2}}^{(0)} -\sum_{\ell_{1}=1}^{\ell-1}\sum_{\ell_{2}=1}^{\ell_{1}-1}     a_{2\ell-2\ell_{1}}^{(0)}     a_{2\ell_{1}-2\ell_{2}}^{(0)}     a_{2\ell_{2}}^{(0)}\nonumber
\end{align}
for $\ell \geq 1$. This has been simplified by noting that $a_{2\ell}^{(0)}$ is imaginary for all $\ell$. Aside from the $c_n$ factors, this is the same recurrence equation for the inner equation in the matched asymptotic expansion approach \cite{lustri2025exponential}. 

To calculate the exponential coefficients, we substitute the expansion \eqref{eq:simpler-ansatz} into (\ref{e:innereq3}) and balance powers of $\eta$ and exponential factors, giving
\begin{equation}
\sum_{n=1}^{\infty}c_{n}     \sum_{m=0}^{2n}     \left(-A_{i}\right)^{2n-m}\frac{(2n)!}{m!     (2n-m)!}     \diff{^{m}\Phi^{(i)}}{\eta^{m}}+3\left(\Phi^{(0)}\right)^{2}\Phi^{(i)}=0     .\label{eq:ODE-for-exp-sectors}
\end{equation}
The Laurent series for $\Phi^{(i)}$ has the form given in \eqref{eq:transs-pert-ansatz} with $\nu_i=1$. This is a linear equation for $\Phi^{(i)}$, and thus we can obtain an algebraic equation for the possible singulants from \eqref{eq:ODE-for-exp-sectors} using the series solution
for $\Phi^{(0)}$ from \eqref{eq:transs-pert-ansatz} and the ansatz \eqref{eq:transs-pert-ansatz} for $\Phi^{(i)}$. This gives the same equation as that for $\chi'$ from \eqref{e:chifam}, which we expect if $\chi$ is linear (ie. $\chi'$ is constant). This equivalence permits us to identify values of $A$ with values of $\chi'$, and therefore the Stokes constants in the inner region with the values of $\Lambda(\mu)$ in the late-order terms. We will let $A_1 = 2\pi \i$, and when the distinction between key singulants is necessary, they will be denoted by $A_{1,\pm} =  \pm 2 \pi \i$ and $A_{2,\pm} = \chi_{2,\pm}'$.

Applying \eqref{eq:transs-pert-ansatz} to (\ref{eq:ODE-for-exp-sectors}) and balancing coefficients of $\eta^{-\ell}$ gives for $\ell \geq 0$
\begin{equation}
\sum_{m=1}^{\ell}a_{\ell-m}^{(i)}\frac{\Gamma(\nu_{i}+\ell)}{\Gamma(m+1)     \Gamma(\nu_{i}+\ell-m)}     \frac{d^{m}F(A_{i})}{dA_{i}^{m}}=0,
\end{equation}
with  $F\left(A_i\right) =2(\cosh A_{i}-1)(1+2\mu     (\cosh A_{i}+1))$.
For $\ell \geq 0$, the coefficients solve
\begin{align}\label{eq:coeff-transs-exp-equation}
F'(A_{i})(\nu_{i}+\ell+1)a_{\ell+1}^{(i)}=-\sum_{m=0}^{\ell}a_{\ell-m}^{(i)}\frac{\Gamma(\nu_{i}+\ell+2)}{\Gamma(m+3)\Gamma(\nu_{i}+\ell-m)}F^{(m+2)}(&A_{i})\\-3\sum_{r=0}^{\left\lfloor \ell/2\right\rfloor }\sum_{\ell_{1}=0}^{r}a_{\ell_{1}}^{(i)}a_{r-\ell_{1}}^{(i)}&a_{\ell-2r}^{(i)},\nonumber
\end{align}

For the singulant $A_{2,\pm}$, we find $\beta_{i}=0$. We are free to set $a_{0}^{(i)}=1$ without loss of generality, as changing $a_0^{i}$ simply alters the value of $\sigma_i$. Now (\ref{eq:coeff-transs-exp-equation}) can be solved for all $\ell\ge0$. Note that $F'(A_{2,\pm})\ne0$.

For the singulant $A_{1,\pm}$ we have $F'(A_{i})=0$, and the equation for $\ell=0$ is identically solved. This is consistent with the observation that the prefactor equation \eqref{eq:vaporiseQ} in the outer region is also identically solved. We obtain $\nu_{i}=2$ or $\nu_{i}=-3$, and we will focus on the latter in order to match with the outer problem. Solving (\ref{eq:coeff-transs-exp-equation})
for $\ell\ge1$ directly shows that $a_{\ell}^{(2,+)}=(-1)^{\ell}     a_{\ell}^{(2,-)}$ and $a_{\ell}^{(1,+)}=(-1)^{\ell}     a_{\ell}^{(1,-)}$.

\subsection{Asymptotic analysis of the algebraic series}

\begin{figure}
\centering
\subfloat[Borel transform Pad\'{e} poles]{
\includegraphics[width=0.35\textwidth]{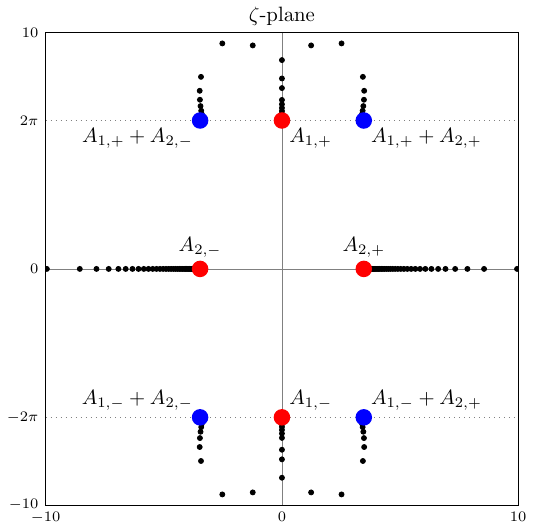}
}\hspace{20pt}
\subfloat[Mapped transform Pad\'{e} poles]{
\includegraphics[width=0.35\textwidth]{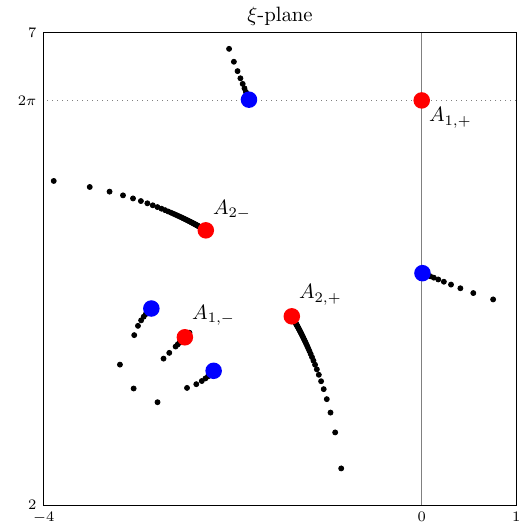}
}
\caption{(a) Singularity structure of the Borel transform of the algebraic series solution for $\mu=-1/30$, determined from the poles of its diagonal Pad\'{e} approximant of order $150$. Branch points corresponding to $A_{2,\pm}$ and $A_{1,\pm}$ are shown in red. Composite singularities are shown in blue. (a) Singularity structure of the conformally-mapped Borel transform with $\mu=-1/30$ around the value of $A_{1,+}=2\pi\mathrm{i}$ (shown in red), determined from the poles of its diagonal Pad\'{e} approximant of order $N=150$.}\label{fig:Borel-pade-sing-1}

\end{figure}

The divergent series $\Phi^{(0)}$ is asymptotic, and we can determine its Borel transform 
\begin{equation}
\mathcal{B}\left[\Phi^{(0)}\right](\zeta)=\sum_{\ell=0}^{\infty}\frac{a_{2\ell}^{(0)}}{\Gamma(\nu_0+2\ell)}     \zeta^{2\ell},
\end{equation}
which has a finite nonzero radius of convergence, as the factorial divergence of the series terms is suppressed. To analyse its singularity structure, we use Pad\'{e} approximation \cite{graves1981pade} to calculate a rational function which approximates the Borel transform and then compute the zeros of its denominator. The result is shown in
Figure \ref{fig:Borel-pade-sing-1}(a) for $\mu=-{1}/{30}$. The leading singularities are at $A_{2,\pm}$ on the real axis and $A_{1,\pm}$ on the imaginary axis. The solution contains an accumulation of poles which approach the leading singularities; this occurs because the Borel transform contains branch cuts, which appear as pole accumulations in rational function approximations \cite{stahl1997convergence}.

It is known that a formal transseries given by \eqref{eq:simpler-ansatz} 
contains asymptotic contributions to the coefficients satisfying
\begin{equation}
a_{\ell}^{(0)}\sim\frac{S_{1}}{2\pi\mathrm{i} A_1^{\ell+\nu_{0}-\nu_{1}}}  \sum_{k=0}^{\infty}     {\Gamma(\ell+\nu_{0}-\nu_{1}-k)}  a_{k}^{(1)}     A_1^{k}    \quad \mathrm{as} \quad \ell \to \infty ,\label{eq:large-order-simple}
\end{equation}
where $S_{1}$ is the so-called Stokes constant \cite{aniceto2019primer,Dingle}. This constant plays a crucial role in Stokes' phenomenon: across a Stokes line, the parameter $\sigma_i$ jumps by $\sigma_i\mapsto\sigma_i+S_{i}$. This can be seen from the behavior of the Borel transform close to the singularity $A_1$: If we multiply the base series by $x^{\nu_{1}}$, then the Borel transform becomes:
\begin{equation}
\mathcal{B}\left[x^{\nu_{1}}\Phi^{(0)}\right](\zeta)=\sum_{\ell=0}^{\infty}\frac{a_{\ell}^{(0)}}{\Gamma(\ell+\nu_{0}-\nu_{1})}     \zeta^{\ell+\nu_{0}-\nu_{1}-1}     ,
\end{equation}
thus effectively removing the factorial growth associated with $A_1$ in the asymptotics. This evaluates to give a log-cut singularity starting at $\zeta=A_1$ in the Borel plane:
\begin{equation}
\left.\mathcal{B}\left[x^{\nu_{1}}\Phi^{(0)}\right](\zeta)\right|_{A_1}=-\frac{S_{1}}{2\pi\mathrm{i}}\mathcal{B}\left[x^{\nu_{1}}\Phi^{(1)}\right]\left(\zeta-A_1\right)     \ln\left(\zeta-A_1\right)+\mathrm{regular\:part}     .\label{eq:Borel-sing-simple}
\end{equation}
In our problem, each of the leading singularities introduces contributions of the form \eqref{eq:large-order-simple},
Using the properties of the even and odd coefficients and the fact that $a_{2\ell+1}^{(0)}=0$ we determine that $S_{1,+}=-S_{1,-}$ and $S_{2,+}=-S_{2,-}$. Thus 
\begin{align}
a_{2\ell}^{(0)}\sim  \frac{2S_{2,+}}{2\pi\mathrm{i}A_{2,+}^{2\ell+1}}    \sum_{k=0}^{\infty}     &\Gamma(2\ell+1-k)    a_{k}^{(2,+)}     A_{2,+}^{k} \\
 & +\frac{2S_{1,+}}{2\pi\mathrm{i}A_{1,+}^{2\ell+4}}     \sum_{k=0}^{\infty}   \Gamma(2\ell+4-k)     a_{k}^{(1,+)}     A_{1,+}^{k}+\cdots,   \nonumber
\end{align}
where $a_{\ell}^{(i)}$ were obtained above, $\nu_{0}=1$, $\nu_{2,\pm}=0$, $\nu_{1,\pm}=-3$, $A_{2,\pm}=\chi_{2,\pm}'$
and $A_{1,\pm}=\pm2\pi\mathrm{i}$. $S_{1,\pm}$ and $S_{2,\pm}$ are the Stokes constants in the
asymptotic behavior.

\subsection{From log cut to simple pole}

If we take a transseries of the type (\ref{eq:simpler-ansatz}) with asymptotic large-order behavior like (\ref{eq:large-order-simple}) and multiply the algebraic series by a factor $x^{\nu_{i}}$ such that the Borel transform removes the exact factorial growth seen in the series terms, then we obtain logarithmic singularities in the Borel plane at $\zeta=A_i$ (\ref{eq:Borel-sing-simple}). This is the origin of the branch cuts seen in Figure \ref{fig:Borel-pade-sing-1}(a).

By selecting a different power of $\eta$, we adjust the order of the branch point in the Borel
plane \cite{aniceto2019large}. If we instead multiply by $\eta^{-\frac{1}{2}+\nu_{1}}$, then the Borel transform,  
\begin{equation}
\mathcal{B}\left[\eta^{-\frac{1}{2}+\nu_{1}}\Phi^{(0)}\right](\zeta)=\sum_{\ell=0}^{\infty}\frac{a_{\ell}^{(0)}}{\Gamma(\ell+\nu_{0}-\nu_{1}+\frac{1}{2})}     \zeta^{\ell+\nu_{0}-\nu_{1}-\frac{1}{2}},\label{eq:halfborel}
\end{equation}
 has square root singularities at $\zeta=A_i$ \cite{aniceto2019primer, aniceto2019large}. In particular,
\begin{equation}
\left.\mathcal{B}\left[\eta^{-\frac{1}{2}+\nu_{1}}\Phi^{(0)}\right](\zeta)\right|_{A_1}=-\frac{S_{1}}{2}     \mathcal{B}\left[x^{-\frac{1}{2}+\nu_{1}}\Phi^{(1)}\right](\zeta-A_1)     ,
\end{equation}
where 
\begin{equation}
\mathcal{B}\left[\eta^{-\frac{1}{2}+\nu_{1}}\Phi^{(1)}\right](\zeta)=\zeta^{-\frac{1}{2}}     \sum_{\ell=0}^{\infty}\frac{a_{\ell}^{(1)}}{\Gamma(\ell+\frac{1}{2})}     \zeta^{\ell}     .
\end{equation}
The leading behavior at the branch point singularity is then given
by 
\begin{equation}
\left.\mathcal{B}\left[\eta^{-\frac{1}{2}+\nu_{1}}\Phi^{(0)}\right](\zeta)\right|_{A_1}=-\frac{S_{1}}{2}     \frac{1}{\sqrt{\zeta-A_1}}\left(\frac{a_{0}^{(1)}}{\Gamma(\frac{1}{2})}+\cdots\right).
\end{equation}
The omitted terms include regular powers of $(\zeta-A_1)$ and their coefficients. 

This modified Borel transform has converted the log-cut singularity at $A_1$ into a square root branch cut. We can now perform a conformal map to the Borel
transform to convert the branch point into a simple pole so that we can extract information from the Borel transform using residue calculus \cite{aniceto2019large}. We perform the change of variable
\begin{equation}
\zeta=A_1-\left(\xi-A_1\right)^{2}     .\label{eq:borel-conformal-map}
\end{equation}
The point $\zeta=A_1$ remains at $\xi=A_1$, or $\zeta = 2 \pi \i$. The Borel transform becomes
\begin{equation}
\left.\mathcal{B}\left[\eta^{-\frac{1}{2}+\beta_{1}}\Phi^{(0)}\right](\xi)\right|_{A_1}=-\frac{S_{1}}{2\mathrm{i}}     \frac{1}{\xi-A_1}\left(\frac{a_{0}^{(1)}}{\Gamma(\frac{1}{2})}+\cdots\right).\label{eq:simplepole}
\end{equation}
The extra regular terms are related to the other coefficients of $\Phi^{(1)}$.

\subsection{Residue calculation of the Stokes constant}

The singular behavior of the Borel transform in \eqref{eq:simplepole} now allows us to determine the Stokes constant $S_{1}$ using residue calculus. Starting from the asymptotic expansion $\Phi^{(0)}$, we calculate the Borel transform \eqref{eq:halfborel} and perform the change of variables \eqref{eq:borel-conformal-map}. We then expand the result at $\xi=A_1-\sqrt{A_1}$ and obtain a rational approximation using the Pad\'{e} approximant centered at this point. This Pad\'{e} approximant will contain an isolated simple pole at $\xi=A_1$. The Stokes constant is given by 
\begin{equation}
S_{1}=\frac{2\mathrm{i}\Gamma(\frac{1}{2})}{a_{0}^{(0)}}     \mathrm{Res}_{\xi=A_1}\mathrm{BP}_{N}\left[\eta^{-\frac{1}{2}+\nu_{1}}\Phi^{(0)}\right]\left(\xi\right)     ,
\end{equation}
where $\mathrm{BP}_{N}$ denotes the diagonal Pad\'{e} approximant of order $N$ of the conformally-mapped Borel transform (for which we need at least $2N$ terms), around the point $\xi=A_1-\sqrt{A_1}$. It can be calculated using standard techniques from \cite{graves1981pade}.

\begin{figure}
    \centering
    \includegraphics[width=0.65\textwidth]{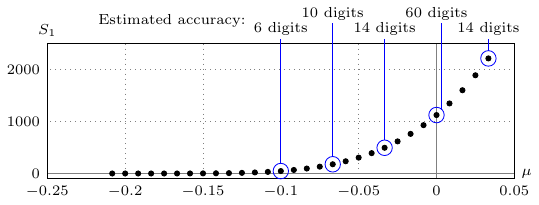}
    \caption{Estimates of $S_1$ obtained for different values of $\mu$ with $N = 300$, including some positive values. The estimated accuracy is shown for representative cases, obtained by comparing $N= 300$ to $N = 400$ and determining how many decimal places remained constant. As $|\mu|$ increases, the numerical accuracy of $S_1$ is reduced, but is still sufficient for this analysis. At $\mu = 0$ the computation is highly accurate, as the contribution from $\chi_2$ vanishes and does not interfere with the computation of $S_1$.}
    \label{fig:StokesConstants}
\end{figure}

In Figure \ref{fig:Borel-pade-sing-1}(b) we show the singularities of the Pad\'{e} approximant of the conformal Borel transform for $\mu=-1/30$, with the choice of singulant $A_1=2\pi\mathrm{i}$ (and $\nu_1=-3$). The simple pole at $\xi=2\pi\i$ is apparent in this figure. In Figure \ref{fig:StokesConstants}, we determine the value of $S_1$ for a range of $\mu$. These results are consistent with \cite{lustri2025exponential} for $\mu = 0$, and show that $S_1$ decays rapidly as $\mu$ grows in the negative direction.

This procedure can be performed to calculate the Stokes multiplier of any leading singularity; however, the analytic approximation of the Borel transform given by the Pad\'{e} approximant depends on the number of terms that have been determined of the leading series, $\Phi^{(0)}$.  While the Stokes multipliers of the singularities closest to the origin of the Borel plane can be calculated to high accuracy, for singularities further from the origin (such as the singularities $A_{1,\pm}$, which are further from the origin than $A_{2,\pm}$) the accuracy is dependent on how many singularities are closer to the origin than the singularity of interest, and how much closer to the origin they are. 

The final step is to match the Stokes constant $S_1$ with the constant $\Lambda(\mu)$. By writing the $k = 0$ term of the late-order asymptotics of $\hat{q}_{j}$, using $a_{\ell}^{(0)}$ with $2j =  \ell$ in \eqref{eq:large-order-simple}, in terms of the outer variable $\tilde{z}$ and requiring it to be consistent with the inner limit of the late-order ansatz for $q_j$ as $\tilde{z} \to \tilde{z}_s$ \eqref{e:localansatz}, it can be seen directly that 
\begin{equation}
   \frac{S_{1}\Gamma(2j+\nu_0 - \nu_1)}{2\pi\mathrm{i} A_1^{2j+\nu_0 - \nu_1}}   = \frac{\Lambda(\mu) \Gamma(2j + \gamma)}{(2\pi {\rm i})^{2j+\gamma}}.
\end{equation}
Using $A_1 = 2 \pi \i$, $\nu_0 = 1$, $\nu_1 = -3$ and $\gamma = 4$ gives $ 2 \pi \i\Lambda(\mu) = S_1$, which means we have fully determined $q_{\mathrm{exp}}$ from \eqref{e:remainder1}.

\section{Stability}\label{S:Stability}

In Sections \ref{S:stokesphen} and \ref{S:expcont} we found that as $\eps \to  0$, if $z > z_0$ then
\begin{equation}
q(\tilde{z}) \sim \sum_{j=0}^{N-1} \eps^{2j} q_j(\tilde{z})  + \frac{2\pi}{\eps^4} U e^{-\beta\pi^2/\eps}\sin\left(\frac{2\pi \tilde{z}}{\eps}\right).
\end{equation}
We will subsequently refer to this steady solution as $q_s(\tilde{z})$. It satisfies
\begin{equation}
q_s(z) \sim 2\sqrt{2}e^{-\tilde{z}/\beta} + \sum_{j=1}^{N-1} \eps^{2j} q_j(z)  - \frac{2\pi}{\eps^4} \left(\frac{5 \beta^3 {\rm i}\Lambda}{4}  \mathrm{e}^{\tilde{z}/\beta}\right)   e^{-\beta\pi^2/\eps}\sin\left(\frac{2\pi \tilde{z}}{\eps}\right),
\end{equation}
as ${\tilde{z}}\to \infty$, $\eps \to 0$. We assume that $n_0$, and hence $z_0$, is chosen to give either a site-centered or inter-site-centered solution. We introduce perturbations of the form
\begin{equation}
Q_n = e^{{\rm i} t}\left[q_s(\tilde{z}) + f(\tilde{z}) e^{\lambda t} +g^*(\tilde{z}) e^{\lambda^* t}\right].
\end{equation}
We set $v = f+g$, $w = f-g$ and obtain by direct substitution that
\begin{align}\label{e:lambdaeq1}
\mathcal{L}_0 v = -{\rm i}\lambda w,\qquad
\mathcal{L}_1 w = {\rm i}\lambda v,
\end{align}
where
\begin{align}
\mathcal{L}_j &= 2\sum_{m=1}^{\infty}\frac{c_m\eps^{2m-2}}{(2m)!}\frac{d^{2m}}{d{\tilde{z}}^{2m}} + \kappa_j q_s^2  - 1,
\end{align}
with $\kappa_0 = 3$ and $\kappa_1 = 1$. We expand $v$ and $w$ as power series in the eigenvalue $\lambda$ in the limit $\lambda \to 0$, giving the leading-order equations $\mathcal{L}_0 v_0 = 0$ and $\mathcal{L}_1 w_0= 0$.
Given our interest in the eigenvalue associated with the 
translational invariance,
we consider the translation eigenfunction using the well-known solution 
\begin{equation}
\label{e:translate}v_0  = q_s'(\tilde{z}), \qquad w_0 = 0.
\end{equation}

\subsection{Calculating the eigenvalue}

To calculate the eigenvalue associated with translation invariance, we apply \eqref{e:translate} to obtain $w_0 = 0$ and
\begin{equation}
v_0 \sim - \frac{5 |\Lambda|\beta^3 \pi^2}{\eps^5} e^{\tilde{z}/\beta}\mathrm{e}^{-\beta \pi^2/\eps} \cos\left(\frac{2\pi\tilde{z}}{\eps}\right)\quad \mathrm{as} \quad \tilde{z} \to \infty,   \,   \eps \to 0.
\end{equation}
We now balance terms in \eqref{e:lambdaeq1} at $\mathcal{O}(\lambda)$ to obtain $\mathcal{L}_0 v_1 = 0$ and $ \mathcal{L}_1 w_1 = -{\rm i}q_s'(\tilde{z})$. 
The first equation gives $v_1 = 0$. Using $q_s \sim q_0$ as $\eps \to 0$, the second equation gives
\begin{equation}
\label{e:w1b}\beta^2\frac{d^{2}w_1}{dz^{2}}  + \left(2       \sech^2\left(\frac{\tilde{z}}{\beta}\right)  - 1\right) w_1 \sim -\frac{{\rm i}\sqrt{2}}{\beta}     \sech\left(\frac{\tilde{z}}{\beta}\right)\tanh\left(\frac{\tilde{z}}{\beta}\right).
\end{equation}
Solving this using variation of parameters, as in \cite{lustri2025exponential}, gives 
 \begin{equation}
 w_1= \frac{{\rm i}\sqrt{2}}{2 \beta^2} \left(\tilde{z}+\alpha\right)\sech\left(\frac{\tilde{z}}{\beta}\right),
 \end{equation}
 where $\alpha$ is an arbitrary constant. As $w_1(z)$ decays in both directions, we do not obtain a condition on the choice of $\alpha$. 
 Balancing terms in \eqref{e:lambdaeq1} at $\mathcal{O}(\lambda^2)$ gives $\mathcal{L}_0 v_2 = -{\rm i}w_1$ and $\mathcal{L}_1 w_2 = 0$. 
We see that $w_2 = 0$, and $v_2$ satisfies
\begin{equation}
\beta^2 \frac{d^{2}v_2}{d\tilde{z}^{2}} + (3u_s^2 -1)  v_2 = -\frac{\sqrt{2}}{2\beta^2}(\tilde{z} + \alpha)\,\sech\left(\frac{\tilde{z}}{\beta}\right).\label{e:inhom}
\end{equation}
The homogeneous solutions to this equation were given previously in \eqref{e:U2}. 
Solving \eqref{e:inhom} with the condition that the solution decays as $\tilde{z} \to -\infty$ gives
\begin{align}\label{e:v1c}
v_2 =& \frac{1}{2\sqrt{2}\beta}U_1(\tilde{z})\int_a^{\tilde{z}} U_2(s)(s + \alpha)\,\sech\left(\frac{s}{\beta}\right)ds\\&- \frac{1}{2\sqrt{2}\beta}U_2({\tilde{z}})\int_{-\infty}^{\tilde{z}} U_1(s)(s + \alpha)\,\sech\left(\frac{s}{\beta}\right) ds
\sim -\frac{e^{{\tilde{z}/\beta}}}{4\sqrt{2}\beta}\nonumber
\end{align}
as $\tilde{z} \to \infty$. We can now predict the behavior of the tail. On lattice points,
\begin{equation}
v \sim v_0 + \lambda^2 v_2 \sim - \frac{5 |\Lambda(\mu)|\beta^3\pi^2}{\eps^5} e^{\tilde{z}/\beta} e^{-\beta\pi^2/\eps} \cos\left(\frac{2\pi\tilde{z}}{\eps}\right) - \frac{\lambda^2 e^{\tilde{z}/\beta}}{4\sqrt{2}\beta},
\end{equation} 
as $\tilde{z} \to \infty$ and $\eps \to 0$. This differs from the result in \cite{lustri2025exponential} only by the presence of $\beta$. As both terms grow large in the limit $z \to \infty$, requiring them to precisely cancel gives the condition that 
$v_0 \sim -\lambda^2 v_2$ as $\eps \to 0$. Solving this expression for $\lambda$ and writing the result in terms of the original discrete variable $n$ gives
\begin{align}
\lambda^2 \sim -\frac{2 0\sqrt{2}|\Lambda(\mu)|\beta^4\pi^2}{\eps^{5}} e^{-\beta\pi^2/\eps}\cos\left({2\pi(n-n_0)}\right)\quad \mathrm{as} \quad \eps \to 0.
\end{align}
For on-site pinning  $n_0$ is an integer, so $\lambda^2 < 0$. The eigenvalues are imaginary, and the on-site standing wave is stable. For inter-site pinning $n_0$ is a half-integer, so there is a positive real value for $\lambda$ and the corresponding waveform is unstable. In both cases, the eigenvalue has magnitude 
\begin{align}|\lambda|\sim \frac{2^{5/4}5^{1/2}\beta^2\pi|\Lambda(\mu)|^{1/2} e^{-\beta\pi^2/(2\eps)}}{\eps^{5/2}} \quad \mathrm{as} \quad \eps \to 0.
\label{original}
\end{align}
To later validate this prediction numerically, 
we seek the first correction term to $\lambda$.

\subsection{Correction Term}

To determine the correction to the eigenvalue, we must match the asymptotic solution for $q$ with two terms of an inner expansion in the neighborhood of the singularity at $\tilde{z} = \tilde{z}_s$. As in \cite{lustri2025exponential}, we utilise the method from \cite{king_chapman_2001} to find the first correction to the pole locations in $q(\tilde{z})$.

Combining the local expansions from \eqref{e:zsnnn} and \eqref{e:q1nearzs} near $\tilde{z} = \tilde{z}_s$ gives
\begin{equation}
q \sim -\frac{{\rm i}\beta\sqrt{2}}{\tilde{z}-\tilde{z}_s} + \frac{{\rm i}(\tilde{z}-\tilde{z}_s)}{3\beta{\sqrt{2}}} + \eps^2\left(\frac{{\rm i}\sqrt{2}(1+16\mu)}{3\beta(\tilde{z}-\tilde{z}_s)^3} + \frac{\pi(1+16\mu)}{24\beta^2\sqrt{2}(\tilde{z}-\tilde{z}_s)^2}\right),
\end{equation}
as $\eps \to 0$ and $\tilde{z} \to \tilde{z}_s$. 
Selecting the inner variables $\eps \eta = \tilde{z}-\tilde{z}_s$ and $\hat{q}(\eta) = \eps q(\tilde{z})$ gives
\begin{equation}
\hat{q}(\eta) \sim \left(-\frac{{\rm i}\beta\sqrt{2}}{\eta} + \frac{{\rm i}\sqrt{2}}{3\beta\eta^3}\right) + \frac{\eps\pi(1+16\mu)}{24\beta^2\sqrt{2}\eta} + \mathcal{O}(\eps^2)
\end{equation}
as $\eps \to 0$. 
We define the alternative inner variable 
$\eps \hat{\eta} =(\tilde{z} - \tilde{z}_{s} )- \tfrac{\pi {\rm i}}{48\beta^3}(1+16\mu)\eps^2$, 
which is chosen to precisely cancel the $\mathcal{O}(\eps)$ correction term. This leaves
\begin{equation}
\hat{q}(\eta) \sim \left(-\frac{{\rm i}\beta\sqrt{2}}{\hat{\eta}} + \frac{{\rm i}\sqrt{2}}{3\beta\hat{\eta}^3}\right) + \mathcal{O}(\eps^2)
\end{equation}
as $\eps \to 0$, which is accurate to the required two terms. This tells us that the true pole location has an $\mathcal{O}(\eps^2)$ correction, and we set 
$\tilde{z}_{s} = \tfrac{{\rm i}\beta\pi}{2}+ \tfrac{\pi {\rm i}}{48\beta^3}(1+16\mu)\eps^2$. 
This produces the adjusted exponentially-small remainder as $\eps \to 0$ and $\tilde{z} \to \infty$
\begin{align}\label{e:remainder1st}
R &\sim -\frac{5\pi\beta^3|\Lambda(\mu)|}{2\eps^4} e^{\tilde{z}} e^{-\beta\pi^2/\eps} e^{- \eps\pi^2(1+16\mu)/(24\beta^3)}\sin\left(\frac{2\pi \tilde{z}}{\eps}\right).
\end{align}
We can now find the first correction to the eigenvalue $\lambda$. Equation \eqref{original} becomes
\begin{align}
|\lambda|&
\sim \frac{2^{5/4}5^{1/2}\pi\beta^2|\Lambda(\mu)|^{1/2} e^{-\beta\pi^2/(2\eps)}}{\eps^{5/2}} \left(1 - \frac{\eps\pi^2(1+16\mu)}{48\beta^3}\right)\quad \mathrm{as} \quad \eps \to 0,\label{eq:result}
\end{align}
where we have only retained the first correction term in $\eps$.

\section{Numerical Comparison}\label{S:numerics}

\begin{figure}
\centering
\includegraphics[width=0.99\textwidth]{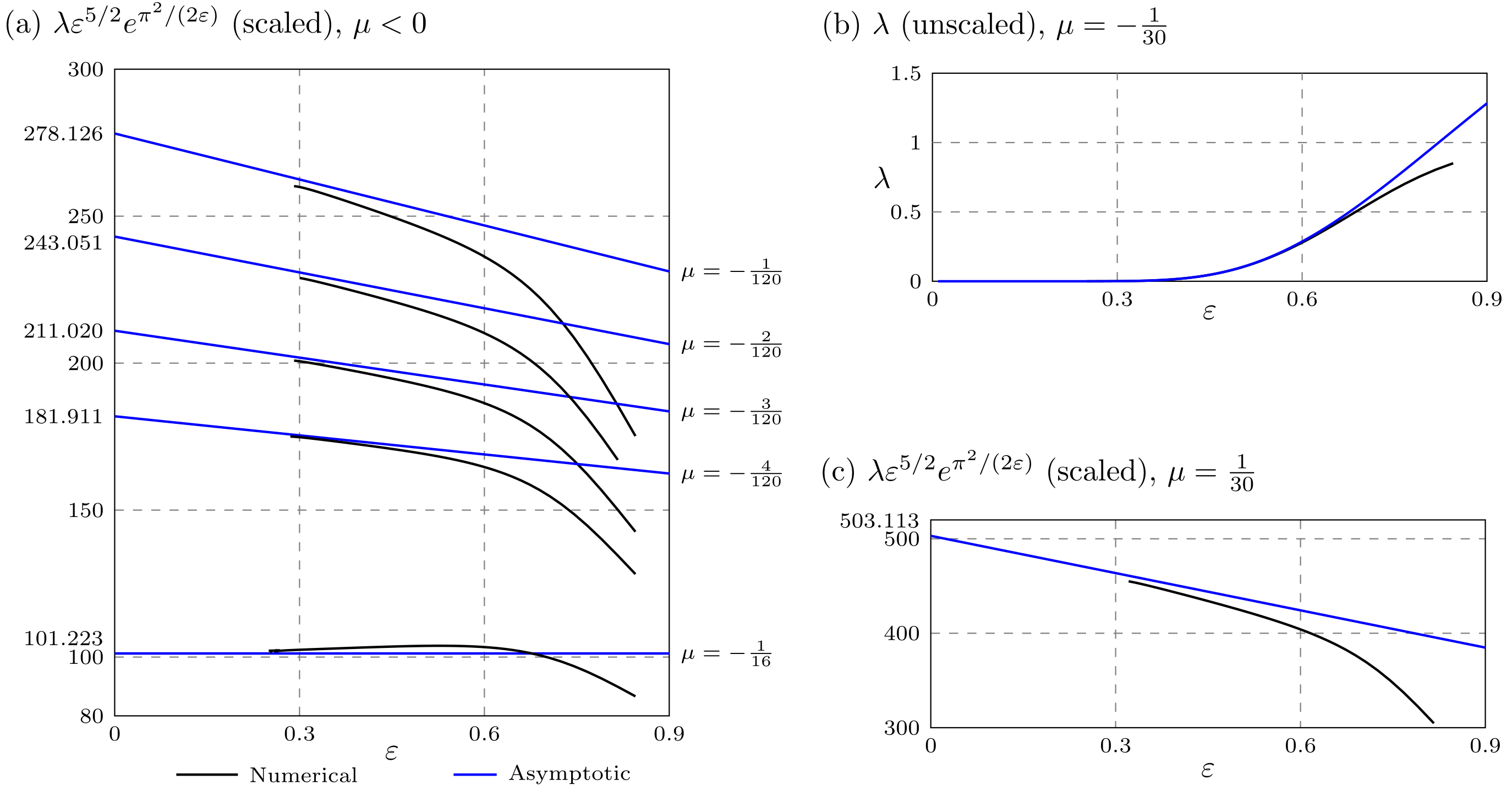}
\caption{Asymptotic vs Numerical computation of
the relevant exponentially small imaginary eigenvalue associated with translation in the case of the site-centered standing wave
of the model with (NN and) NNN interactions for different values of $\mu$. Numerical simulations are black, and asymptotic predictions are blue.}\label{fig:num}
\end{figure}

In order to examine the validity of the original
prediction of \eqref{eq:result}, we have computed the eigenvalues over a range of $\mu$ for site-centered solitary waves and their associated stability. 
Similar results can be obtained for the inter-site-centered
solitary waves and their respective real eigenvalues.
These computations use existing numerical methods, such as those from~\cite{johaub,Todd_Kapitula_2001,kev09}; Newton-type
iterations allow for the calculation of the stationary solution
and a standard eigenvalue solver such as {\tt eigs}
(within Matlab) enables the
computation of the corresponding eigenvalues.
The distinguishing feature of this analysis is that we push further towards the small-$\varepsilon$ limit than has been previously considered. 

A plot of $\lambda$ against $\eps$ for $\mu = -{1}/{30}$ is presented in Figure \ref{fig:num}(b). It  shows good qualitative match for the 
dependence of the relevant eigenvalue on $\eps$. Despite this, Figure \ref{fig:num}(b) alone is not sufficient to determine whether the asymptotic analysis is accurate, as the visible behavior is dominated by the exponential term. Instead, we study a scaled version of this plot in Figure \ref{fig:num}(a).

The scaled eigenvalue comparison is obtained through multiplying both the numerical results and asymptotic predictions by $\eps^{5/2} \mathrm{e}^{\beta\pi^2/(2\eps)}$, which removes the $\eps$-dependence from the leading-order scaling in \eqref{eq:result}. This rescaling imposes a multiplication by a {\it growing exponential},
providing an extremely sensitive diagnostic 
for
potential error in 
the 
exponent. This
clearly 
confirms not only the accuracy of the exponential dependence  
of 
the power law prefactor; it also showcases the 
approach to our asymptotic prediction in a quite definitive manner. 

Figure \ref{fig:num}(a) shows clear agreement between
the asymptotic prediction of \eqref{eq:result} (blue
lines) and the full numerical results (black lines). This comparison is challenging, as  the variation of the relevant eigenvalue occurs over more than 6 orders of magnitude for the interval shown.
Confirming the predictions for smaller values of $\eps$ would require tools that are considerably more refined at the level of the numerical eigenvalue solver which, while not impossible, are outside the scope of the present work.

Note that the results in \ref{fig:num}(a) also include the significant parameter choice $\mu = -{1}/{16}$, which corresponds to the finite-difference case. It also happens to be the case where the first correction vanishes entirely. Interestingly, this feature is 
appropriately reflected in the relevant 
numerical results; see the lower-most curve in Figure \ref{fig:num}(a).

\section{Conclusions and Future Directions}\label{S:conclusions}

In this study, we considered localized solitary wave solutions to the discrete nonlinear Schr\"{o}dinger lattice equation with the addition of next-to-nearest neighbor interactions. We found that, for $\mu > -1/4$, the lattice supports on-site and inter-site solutions and that only the on-site solutions are stable, in line with what was
previously known for the purely NN case~\cite{johaub,Todd_Kapitula_2001,kev09,campbell}. We also found that no such localized solutions can exist for $\mu < -1/4$. 

Beyond determining the stability of the solutions for $\mu > -1/4$, we were also able to calculate the asymptotic dependence of the exponentially small eigenvalues on the lattice spacing $\varepsilon$, which we validated against numerical computations. Notably, the case $\mu = -1/16$ corresponds to the fourth-order finite difference spatial discretization of the continuous nonlinear Schr\"{o}dinger equation.
This led to the disappearance of the leading order correction in our asymptotic analysis; this feature was reflected in the corresponding
numerical eigenvalue computations.

Determining the eigenvalues required the use of exponential asymptotic techniques that extend beyond those commonly seen in the applied mathematics literature, which relies on using matched asymptotic expansions near singular points in the analytic continuation of the solution. The presence of additional small exponential terms in the NNN analysis meant that the eigenvalues could not be determined in this fashion, and instead required us to use conformal Borel-Pad\'{e} analysis from \cite{aniceto2019large} in the large-$|\eta|$ limit to study Stokes' phenomenon in a region near the singularity.

This analysis can also be extended to the case of positive $\mu$
(which is, as explained previously, more physically relevant), and have included one such example in Figure \ref{fig:num}(c) to show that such calculations are possible. As the Borel plane now contains four branch points with $|\chi'| < 2\pi$, the Stokes multiplier calculations have a slightly reduced accuracy compared to negative $\mu$, but they are still accurate to a comparable number of digits. 
Studying this positive $\mu$ case in higher accuracy, as well
as extending the present considerations to waveforms in higher
dimensions and different variants of the DNLS
model (e.g., the saturable case, the so-called Salerno model 
etc.)~\cite{kev09} are a selection of the many interesting 
directions relevant for future study. Such explorations are
currently in progress and will be reported in future publications.

The Borel-Pad\'{e} framework presented here is very general. This approach is more flexible than the standard matched asymptotic expansion method, and associates late-order terms of the parametric outer expansion with the appropriate Stokes' multipliers from the non-parametric inner expansion near singularities. This method can be used to determine both dominant and subdominant Stokes constants in parametric expansions, which was not previously possible, and opens up a new line of approach for future exponential asymptotic studies.

\appendix
\section{Stokes switching}\label{app:stokes}
Truncating \eqref{e:series} after $N$ terms gives
\begin{equation}
q(\tilde{z}) = \sum_{j=0}^{N-1} \eps^{2n}q_n(\tilde{z}) + R(\tilde{z}).
\end{equation}
The optimal truncation is given by $N = \tfrac{|\chi|}{2\eps} + \alpha$, where $\alpha \in [0,1)$ is chosen so that $N \in \mathbb{Z}$. $R(\tilde{z})$ is the remainder after truncation, and we write it in the form $R = \mathcal{S} U\mathrm{e}^{-\chi/\eps}$
where $\mathcal{S}$ behaves as a constant everywhere except in the neighborhood of a Stokes line, across which it changes rapidly in value. This gives, after simplification,
\begin{align}
2\left[\sum_{m=1}^{\infty}\binom{2m}{2}\frac{c_m(\chi')^{2m-2}}{(2m)!}\right]\frac{d^2}{d\tilde{z}^2}\left(\mathcal{S} U\right)\mathrm{e}^{-\chi/\eps}+ (3 u_0^2 - 1) \mathcal{S}U\mathrm{e}^{-\chi/\eps}&\\ \sim -2\sum_{p=N}^{\infty}\eps^{2p}\sum_{m=p-N+2}^{p+1}\frac{c_m}{(2m)!}&\frac{d^{2m}q_{p-m+1}}{d\tilde{z}^{2m}}.\nonumber
\end{align}
Writing $\chi = r\mathrm{e}^{\i\theta}$ and following the same sequence of steps as in \cite{lustri2025exponential} gives
\begin{align}
\beta^2\frac{d^2}{d{\tilde{z}}^2}\left(\mathcal{S} U\right)+ (3q_0^2 - 1) \mathcal{S} U  \sim \frac{2 \mathrm{e}^{-r/\eps - \mathrm{i}\theta(r/\eps + \gamma+2\alpha)}\sqrt{2\pi}U }{\eps^{\gamma+3/2}r^{1/2}}\frac{f_1(\theta)+f_2(\theta)}{1-\mathrm{e}^{-2\mathrm{i}\theta}}\mathrm{e}^{\chi/\eps},
\end{align}
where $f_1(\theta) = \cos(2\pi\mathrm{e}^{-\mathrm{i}\theta}) - 1$ and $f_2(\theta) = \cos(4\pi\mathrm{e}^{-\mathrm{i}\theta}) - 1$. 
The presence of $f_2(\theta)$ is the only substantial difference between this analysis and the analysis from \cite{lustri2025exponential}. 
We therefore follow essentially the same steps as \cite{lustri2025exponential}, as well as the result that $f_1(\eps^{1/2}\vartheta)+f_2(\eps^{1/2}\vartheta) \sim -\mathrm{i}\pi^2\beta^2\eps^{1/2}\vartheta$ as $\eps \to 0$,
to determine that 
\begin{equation}\label{e:remainder4}
q_{\mathrm{exp}} \sim \frac{2\pi}{\eps^4} U e^{-\beta\pi^2/\eps}\sin\left(\frac{2\pi \tilde{z}}{\eps}\right) \quad \mathrm{as} \quad \eps \to 0
\end{equation}
for $\tilde{z} > 0$. There is no exponentially small contribution present for $\tilde{z} < 0$.

\section*{Acknowledgments}
CJL thanks S. J. Chapman for valuable discussions.

\bibliographystyle{siamplain}
\bibliography{DNLS_Refs}

\begin{thebibliography}{10}

\bibitem{cole}
{\sc M.~J. Ablowitz and J.~T. Cole}, {\em Nonlinear optical waveguide lattices:
  Asymptotic analysis, solitons, and topological insulators}, Physica D, 440
  (2022), p.~133440,
  \url{https://doi.org/https://doi.org/10.1016/j.physd.2022.133440}.

\bibitem{aniceto2019primer}
{\sc I.~Aniceto, G.~Ba{\c{s}}ar, and R.~Schiappa}, {\em A primer on resurgent
  transseries and their asymptotics}, Phys. Rep., 809 (2019), pp.~1--135.

\bibitem{aniceto2019large}
{\sc I.~Aniceto, J.~Jankowski, B.~Meiring, and M.~Spali{\'n}ski}, {\em The
  large proper-time expansion of {Y}ang-{M}ills plasma as a resurgent
  transseries}, J. High Energy Phys., 2019 (2019), pp.~1--36.

\bibitem{aniceto2012resurgence}
{\sc I.~Aniceto, R.~Schiappa, and M.~Vonk}, {\em The resurgence of instantons
  in string theory}, Commun. Number Theory Phys., 6 (2012), pp.~339--496.

\bibitem{brazhnyi}
{\sc V.~A. Brazhnyi and V.~V. Konotop}, {\em Theory of nonlinear matter waves
  in optical lattices}, Mod. Phys. Lett. B, 18 (2004), pp.~627--651,
  \url{https://doi.org/10.1142/S0217984904007190}.

\bibitem{caliceti2007useful}
{\sc E.~Caliceti, M.~Meyer-{H}ermann, P.~Ribeca, A.~Surzhykov, and U.~D.
  Jentschura}, {\em From useful algorithms for slowly convergent series to
  physical predictions based on divergent perturbative expansions}, Phys. Rep.,
  446 (2007), pp.~1--96.

\bibitem{Chapman}
{\sc S.~J. Chapman, J.~R. King, and K.~L. Adams}, {\em Exponential asymptotics
  and {S}tokes lines in nonlinear ordinary differential equations}, Proc. Roy.
  Soc. Lond. {A}, 454 (1998), pp.~2733--2755.

\bibitem{Chong2018}
{\sc C.~Chong and P.~G. Kevrekidis}, {\em {Coherent Structures in Granular
  Crystals: From Experiment and Modelling to Computation and Mathematical
  Analysis}}, Springer International Publishing, 2018.

\bibitem{costin2021conformal}
{\sc O.~Costin and G.~V. Dunne}, {\em Conformal and uniformizing maps in
  {B}orel analysis}, The European Physical Journal Special Topics, 230 (2021),
  pp.~2679--2690.

\bibitem{cruickshank2025singlesitemultisitesolitonsbright}
{\sc R.~Cruickshank, F.~Lorenzi, A.~{La Rooij}, E.~Kerr, T.~Hilker, S.~Kuhr,
  L.~Salasnich, and E.~Haller}, {\em Single-site and multi-site solitons of
  bright matter-waves in optical lattices}, 2025,
  \url{https://arxiv.org/abs/2504.11046}.

\bibitem{di2020looking}
{\sc L.~{Di Pietro} and M.~Serone}, {\em Looking through the {QCD} conformal
  window with perturbation theory}, J. High Energy Phys., 2020 (2020),
  pp.~1--35.

\bibitem{Dingle}
{\sc R.~B. Dingle}, {\em Asymptotic Expansions: {T}heir Derivation and
  Interpretation}, Academic Press, New York, NY, USA, 1973.

\bibitem{Dreisow:08}
{\sc F.~Dreisow, A.~Szameit, M.~Heinrich, T.~Pertsch, S.~Nolte, and
  A.~T\"{u}nnermann}, {\em Second-order coupling in femtosecond-laser-written
  waveguide arrays}, Opt. Lett., 33 (2008), pp.~2689--2691,
  \url{https://doi.org/10.1364/OL.33.002689}.

\bibitem{Dreisow:11}
{\sc F.~Dreisow, G.~Wang, M.~Heinrich, R.~Keil, A.~T\"{u}nnermann, S.~Nolte,
  and A.~Szameit}, {\em Observation of anharmonic bloch oscillations}, Opt.
  Lett., 36 (2011), pp.~3963--3965, \url{https://doi.org/10.1364/OL.36.003963}.

\bibitem{Efremidis2002}
{\sc N.~K. Efremidis and D.~N. Christodoulides}, {\em {Discrete solitons in
  nonlinear zigzag optical waveguide arrays with tailored diffraction
  properties}}, Phys. Rev. E, 65 (2002), p.~056607,
  \url{https://doi.org/10.1103/PhysRevE.65.056607}.

\bibitem{chriseil}
{\sc J.~C. Eilbeck and M.~Johansson}, {\em The discrete nonlinear
  {S}chr\"odinger equation - 20 years on}, Localization and Energy Transfer in
  Nonlinear Systems,  (2003), pp.~pp. 44--67.

\bibitem{eynard2023arxiv}
{\sc B.~Eynard, E.~{Garcia-Failde}, P.~Gregori, R.~Schiappa, and D.~Lewanski},
  {\em Resurgent asymptotics of jackiw-teitelboim gravity and the
  nonperturbative topological recursion}, 2025,
  \url{https://arxiv.org/abs/2305.16940}.

\bibitem{graves1981pade}
{\sc P.~Graves-Morris and G.~Baker}, {\em Pad{\'e} approximants, encyclopedia
  of mathematics}, 1981.

\bibitem{heller2022relativistic}
{\sc M.~P. Heller, A.~Serantes, M.~Spali{\'n}ski, V.~Svensson, and B.~Withers},
  {\em Relativistic hydrodynamics: a singulant perspective}, Phys. Rev. X, 12
  (2022), p.~041010.

\bibitem{HU2020126448}
{\sc J.~Hu, S.~Li, Z.~Chen, J.~Lü, B.~Liu, and Y.~Li}, {\em Discrete solitons
  in zigzag waveguide arrays with different types of linear mixing between
  nearest-neighbor and next-nearest-neighbor couplings}, Phys. Lett. A, 384
  (2020), p.~126448,
  \url{https://doi.org/https://doi.org/10.1016/j.physleta.2020.126448}.

\bibitem{jentschura2001improved}
{\sc U.~D. Jentschura and G.~Soff}, {\em Improved conformal mapping of the
  {B}orel plane}, J. Phys. A: Math. Gen., 34 (2001), p.~1451.

\bibitem{johaub}
{\sc M.~Johansson and S.~Aubry}, {\em Growth and decay of discrete nonlinear
  {S}chr{\"o}dinger breathers interacting with internal modes or standing-wave
  phonons}, Phys. Rev. E, 61 (2000), pp.~5864--5879,
  \url{https://doi.org/10.1103/PhysRevE.61.5864}.

\bibitem{Todd_Kapitula_2001}
{\sc T.~Kapitula and P.~G. Kevrekidis}, {\em Stability of waves in discrete
  systems}, Nonlinearity, 14 (2001), p.~533,
  \url{https://doi.org/10.1088/0951-7715/14/3/306}.

\bibitem{kev09}
{\sc P.~G. Kevrekidis}, {\em {The {D}iscrete {N}onlinear {S}chr{\"o}dinger
  {E}quation}}, Springer-Verlag, Heidelberg, 2009,
  \url{https://link.springer.com/book/10.1007/978-3-540-89199-4}.

\bibitem{king_chapman_2001}
{\sc J.~R. King and S.~J. Chapman}, {\em Asymptotics beyond all orders and
  {S}tokes lines in nonlinear differential-difference equations}, Eur. J. Appl.
  Math., 12 (2001), p.~433–463.

\bibitem{campbell}
{\sc Y.~S. Kivshar and D.~K. Campbell}, {\em Peierls-{N}abarro potential
  barrier for highly localized nonlinear modes}, Phys. Rev. E, 48 (1993),
  pp.~3077--3081, \url{https://doi.org/10.1103/PhysRevE.48.3077}.

\bibitem{LEDERER20081}
{\sc F.~Lederer, G.~I. Stegeman, D.~N. Christodoulides, G.~Assanto, M.~Segev,
  and Y.~Silberberg}, {\em Discrete solitons in optics}, Phys. Rep., 463
  (2008), pp.~1--126,
  \url{https://doi.org/https://doi.org/10.1016/j.physrep.2008.04.004}.

\bibitem{lustri2023locating}
{\sc C.~J. Lustri, I.~Aniceto, D.~J. VandenHeuvel, and S.~W. McCue}, {\em
  Locating complex singularities of {B}urgers’ equation using exponential
  asymptotics and transseries}, Proc. Roy. Soc. Lond. A, 479 (2023),
  p.~20230516.

\bibitem{lustri2025exponential}
{\sc C.~J. Lustri, P.~G. Kevrekidis, and S.~J. Chapman}, {\em Exponential
  asymptotics for translational modes in the discrete nonlinear
  {S}chr{\"o}dinger model}, Q. Appl. Math. (In Press),  (2025).

\bibitem{marino2022new}
{\sc M.~Mari{\~n}o, R.~Miravitllas, and T.~Reis}, {\em New renormalons from
  analytic trans-series}, J. High Energy Phys., 2022 (2022), pp.~1--61.

\bibitem{RevModPhys.78.179}
{\sc O.~Morsch and M.~Oberthaler}, {\em Dynamics of {B}ose-{E}instein
  condensates in optical lattices}, Rev. Mod. Phys., 78 (2006), pp.~179--215,
  \url{https://doi.org/10.1103/RevModPhys.78.179}.

\bibitem{Daalhuis}
{\sc A.~B. {Olde Daalhuis}, S.~J. Chapman, J.~R. King, J.~R. Ockendon, and
  R.~H. Tew}, {\em Stokes phenomenon and matched asymptotic expansions}, SIAM
  J. Appl. Math., 55 (1995), pp.~1469--1483.

\bibitem{remoissenet}
{\sc M.~Remoissenet}, {\em Waves Called Solitons}, Springer-Verlag, Berlin,
  1999.

\bibitem{Rothos2025}
{\sc V.~M. Rothos, S.~Anastassiou, and K.~G. Hadjifotinou}, {\em Stationary
  solitons in discrete nonlinear {Schr\"odinger} with non-nearest neighbour
  interactions}, Proc. Roy. Soc. Lond. A, 481 (2025), p.~20240539,
  \url{https://doi.org/10.1098/rspa.2024.0539}.

\bibitem{serone2018lambdaphi4}
{\sc M.~Serone, G.~Spada, and G.~Villadoro}, {\em $\lambda$$\phi^4$
  theory—{P}art {I}. the symmetric phase beyond {NNNNNNNNLO}}, J. High Energy
  Phys., 2018 (2018), pp.~1--35.

\bibitem{stahl1997convergence}
{\sc H.~Stahl}, {\em The convergence of {P}ad{\'e} approximants to functions
  with branch points}, J. Approximation Theory, 91 (1997), pp.~139--204.

\bibitem{Szameit:09}
{\sc A.~Szameit, R.~Keil, F.~Dreisow, M.~Heinrich, T.~Pertsch, S.~Nolte, and
  A.~T\"{u}nnermann}, {\em Observation of discrete solitons in lattices with
  second-order interaction}, Opt. Lett., 34 (2009), pp.~2838--2840,
  \url{https://doi.org/10.1364/OL.34.002838}.

\end{thebibliography}
\end{document}